# Microstructural investigation of the aging behavior of (3Y-TZP)-Al$_2$O$_3$ composites

**Sylvain Deville [‡], Jérôme Chevalier, Christelle Dauvergne, Gilbert Fantozzi**

National Institute of Applied Science, Materials Department (GEMPPM-INSA), Associate Research Unit 5510, 20 av. A. Einstein, 69621 Villeurbanne, France

**José F. Bartolomé, José S. Moya**

Instituto de Ciencia de Materiales de Madrid (ICMM), CSIC, Cantoblanco 28049, Madrid, Spain

**Ramón Torrecillas**

Instituto Nacional del Carbón CSIC, C/ Francisco Pintado Fe, 26 , 33011 Oviedo, Spain

[‡] corresponding author: sylvain.deville@insa-lyon.fr, Tel: +33 4 72 43 63 57, Fax: +33 4 72 43 85 28

**ABSTRACT**

The low temperature autoclave aging behavior of zirconia toughened alumina composites processed by a classical powder mixing processing route was analyzed using atomic force microscopy (AFM), scanning electron microscopy and X-Ray diffraction. The transformation was evaluated in terms of nucleation and growth, assessed by XRD. The time-temperature equivalency of the transformation was used to measure an apparent activation energy of the nucleation stage of the transformation of 78 kJ/mol. The microstructural features influencing the transformation were identified, and the influence of the alumina matrix on the transformation was investigated. Transformation progression grain by grain was observed by AFM. Transformation does not only occur in zirconia agglomerates but also in isolated zirconia grains. The matrix could partially inhibit the transformation. This behavior could be rationalized considering the constraining effect of the alumina matrix, shape strain accommodation arguments and microstructural homogeneity effects.

## 1 INTRODUCTION

Zirconia toughened alumina (ZTA) composites are of raising interest as far as biomedical applications are concerned, as an alternative to biomedical grade alumina and zirconia. Indeed, alumina is somewhat brittle, and recent serious problems have been reported [1,2] with yttria-stabilized zirconia aging degradation used in total hip replacement (THR). Aging is related to the tetragonal to monoclinic (t-m) phase transformation of zirconia [3,4], transformation promoted by mechanical and hydrothermal stresses [3,5,6]. This transformation is accompanied by a 4 vol.% increase and 16% shear, leading to microcracking, surface degradation and eventually grain pop-out. This phenomenon, if not controlled, might become very detrimental for biomedical applications such as THR, where very low surface roughness and long-term stability are required [7]. Using ZTA, improved mechanical properties such as strength and crack propagation resistance might extend the lifetime of orthopedic implants [8] while improving the safety operating range. However, since those ceramics contain zirconia, they are likely to undergo aging. Little attention has been paid so far to this phenomenon in ZTA, and the underlying mechanisms are not yet clear. Some microstructural features influencing the aging behavior have been underlined, though the conclusions remain a bit speculative. Most of the experiments carried out so far were indirect observations of the transformation, i.e. XRD experiments [9-12], neutron powder diffraction [13], Infra Red (IR) spectroscopy [14], mechanical testing [15] or differential thermal analysis (DTA) [6].

Direct observations of the transformation were obtained only by scanning electron microscopy (SEM) [16], where the relief changes could hardly be quantified, and transmission electron microscopy (TEM) [17]. However, though TEM studies have been able to show the selective nature of the transformation and stabilizer reduction, the microstructural and chemical environment in TEM experiments might be questioned In particular, for particles observed in thin foils, the influence of the matrix is greatly reduced, since specimen's thickness is limited to a few nanometers. In addition, grinding of specimens brings important microstructural modifications. Kriven [18] used a 1 MeV HVTEM to simulate bulk conditions and was able to observe stress induced transformation and analyze the crystallography of the transformation. He observed twinning on $(100)_m$ planes, and concluded that either the martensitic theory needed to be modified to take into account bulk effects, or that twins formed subsequently as deformation twins to minimize the shape change of the particle due to transformation. It has since been demonstrated that twins in zirconia were self-accommodated martensitic variants [19]. This study raised the interest for taking into account bulk conditions, as opposed to the behavior of thin foils. Transformation was nonetheless stress-induced, and not autoclave-induced.

Among the identified factors influencing the transformation behavior, the zirconia grain size is of prime importance. By limiting zirconia grain growth, it seems possible restricting the transformation. However, this was identified only for transformation occurring during heat treatment at high temperatures (>1200K) or during cooling after sintering [16], which is different to autoclave aging conditions.

Some of the microstructural features that could influence the autoclave aging behavior of zirconia remain ambiguous. Further understandings of these phenomena require local observation of the transformation at the surface of bulk specimens. Recent developments of atomic force microscopy allow observing the transformation-induced surface relief modification at a nanometer scale [20]. The purpose of this paper is to provide further insights on the aging sensitivity of ZTA and the nucleation and growth nature of the transformation. The inherent limited reliability of XRD raised the interest of local observations. Hence, it is demonstrated how the zirconia particle's transformation can be locally observed by comparing atomic force microscopy (AFM) and scanning electron microscopy (SEM) images, providing new informations about the more influent microstructural features, not accessible by other techniques.

## 2 EXPERIMENTAL

### 2.1 Processing

Samples were processed by a classical powder mixing processing route. A high purity alumina powder α-$Al_2O_3$ >99.9 wt.% (Condea HPA 0.5, Ceralox division, Arizona, USA) was mixed with various amounts of yttria-stabilized zirconia powder (3Y-TZP, Tosoh TZ-3YS, Tosoh corporation, Tokyo, Japan). Samples were sintered in air at 1873 K for two hours. The obtained plates were machined to small bars. These bars were mirror-polished by standard procedures on one side by using diamond slurries and pastes down to 1 μm, reaching mean roughness (Ra) values as low as 2 nm (measured by AFM, after polishing). The residual relief does not interfere with the transformation-induced relief observations.

Samples were thermally etched at a selected temperature, in order to form slight thermal grooves at grain boundaries. Etch experiments for determining the optimal conditions were carried out for 12 min, from 1523 K to 1723 K with heating and cooling rate of 400 K/min. This thermal treatment creates local relief between the grains and allows separating each grain on AFM and SEM images. The choice of the etching temperature and its influence on the aging behavior are discussed in the results section.

### 2.2 Measurements

X-ray diffraction data were obtained with a diffractometer using Cu-$K_\alpha$ radiation. The tetragonal/monoclinic zirconia ratio was determined using the integrated intensity (measuring the area under the diffractometer peaks) of the tetragonal (101) and two monoclinic (111) and (-111) peaks as described by Garvie et al. [21] and then revised by Toroya et al. [22]. The measured monoclinic phase fraction is therefore related to the zirconia phase only, i.e. when a measure of 20 vol.% of monoclinic phase is obtained in a 30 vol.% 3Y-TZP-$Al_2O_3$. This means that 20 vol.% of the 3Y-TZP phase is transformed, that is 6 vol.% of the surface interaction volume of the sample (the X-Ray penetration depth is estimated around 5μm). Diffractograms were obtained from 27° to 33°, at a scan speed of 0.2°/min and a step size of 0.02°. Autoclave treatments up to 80 hrs were performed in the temperature range 383 K-413 K.

SEM images were obtained on polished, thermally etched and gold coated samples by using a high resolution scanning electron microscopy (FEI, XL30 ESEM FEG), used in backscattered electrons imaging mode (BEI).

Samples surface was observed by an atomic force microscope (D3100, Digital Instruments Inc.) in contact mode with an average scanning speed of 10 μm/s. The used oxide-sharpened silicon nitride probes had a nominal spring constant of 0.57 N/m, a tip half angle of 35° and a nominal tip radius of curvature of 20 nm. Experiments were performed in air. Polished and thermally etched samples were just cleaned with ethanol, in order to obtain a very clean surface, free of any residual dirt. The same zones were observed by AFM and SEM.

## 3 RESULTS

### 3.1 Surface preparation

Several temperatures were tested in order to determine the best thermal etching conditions. Considering the expected size of relief due to surface transformation during aging (up to 50 nm), grain boundaries height (GBH) around 10 nm is necessary. If GBH is close to the magnitude of transformation-induced relief, local analysis of the transformation becomes more difficult. Grain boundary width must also be kept small, though it is nonetheless much smaller than the average grain size and therefore does not affect the AFM observations analysis.

Measurement of GBH is shown in Fig 1. For each etch temperature, more than 50 GBH were measured to get a statistically significant average value. Thermal grooves develop because the system reaches an equilibrium configuration determined by the grain boundary and surface energies. The thermal groove is dependent, of course, on the temperature [23] but also on the grains crystallographic orientation relationships to the surface, as well as on the crystallographic relationships of the two adjacent grains of each grain boundary. The selected temperature must reveal the grain boundaries while grains surface remains undisturbed. The variation of GBH with temperature (for a 12 min treatment) is shown in Fig 2. 12 min at 1623 K (heating and cooling rate of 400 K/min) is deemed optimal. This allows getting clear grain boundaries on AFM scans and SEM images, while relief changes due to surface transformation remain clear.

**Origin of AFM-SEM contrast**

To make a direct comparison, the same zones were observed by SEM and AFM. If the two corresponding images are obviously similar, the origin of the contrast on the two types of images is worth further explanations.

As far as SEM is concerned, the image is processed from the backscattered electrons signal and a very large contrast is expected when imaging elements with a great difference in atomic number, such as alumina and zirconia. It is however not possible to see the relief change induced by the transformation. This has been already well documented.

As far as AFM is concerned, the contrast origin is very different. In contact mode, the probe is scanning the surface with a constant applied force and the deflection variation is measured. If the surface is chemically inert, the deflection variation is only related to changes in surface relief; it is possible measuring variations in the nanometer range. In some cases, some interaction between the probe and the surface might occur, interfering with the variation of the relief, so that the interpretation of the obtained informations becomes more difficult. Fortunately, there is no such interaction between the probes used here and the surface of ZTA, so that the AFM scans might be interpreted straightforwardly. Two types of images might then be obtained from the scans: either the *height* image or the *derivate* image (also often called the *error* image). The first one is a topological image; the higher the local surface relief, the brighter the image. Theses images might be seen in 3D for more clarity, relief can be quantified. The derivate image is related to the rate of surface relief evolution. When the relief is changing abruptly, the signal is more important. However, when there is no such change, the derivate signal is very low. This allows seeing similar features on an image, e.g. a self accommodating variant with a constant angle will appear with a constant contrast all along its plane, as it can be seen in Fig 3 and 4 of Ref [20]. However, it does not allow measuring height variations. Depending on the type of information being investigated, either one of these two modes have been used.

**Comparison of the aging behavior with and without thermal etching**

To ensure the thermal etching procedure (used to reveal the microstructure in SEM and AFM observations) did not have a detrimental effect on the aging behavior, experiments were carried out on both unetched and etched samples. The monoclinic phase fraction evolution of two samples processed and polished in the same conditions was measured as a function of the aging treatment time. One of the samples was thermally etched for 12 min at 1623 K. Results are plotted in Fig 3. Though the error range cannot be measured directly, it is usually estimated to lie around 4-5% for monoclinic fraction up to 0.15 (e.g. 5 ± 2 %) and less than 5% for larger fractions where the noise/signal ratio becomes smaller. As it can be seen on the graph, the two curves lie within the measurement error range of XRD analysis.

Considering how short the etching procedure is (12 min at dwell temperature), this result is not surprising. A longer treatment may modify the surface relief more drastically. Longer thermal treatments are also known to relax residual stresses [24]. It can therefore be concluded the samples are not affected by the thermal etching procedure in these conditions, validating the procedure chosen here for AFM and SEM observations.

*3.2    XRD Analysis*

The evolution of the monoclinic phase fraction as a function of autoclaving time at 413 K for different zirconia content is plotted in Fig 4. Transformation is occurring more rapidly when the zirconia content in the composite increases. A fast increase is observed during the first few hours, and the transformation is then slowing down to reach a steady rate. The variation in monoclinic phase fraction was also measured for several aging temperatures (Fig 5), in order to measure the apparent activation energy of the transformation, to make a direct comparison with 3Y-TZP. The same behavior is observed for all the temperatures, and the transformation rate is found to be temperature dependent, as expected from 3Y-TZP previous results [25].

*3.3    Microstructure*

High definition ESEM images (e.g. Fig. 6) were used to determine the alumina and zirconia grain size by the linear intercept method. Values are given in table 1. An average zirconia grain size of 0.5 µm was found, independently of the zirconia content in the sample. Thus, any change of aging behavior cannot be related to 3Y-TZP grain size. It has been shown by precedent studies [26,27] that grain size was the more important factor controlling the aging sensitivity. The point of the analysis performed here was to show that the zirconia grain size is almost independent on the zirconia content in the composite in this case, so this factor will have, if any, a minor influence on the aging behavior of the composite. Hence, differences in the aging behavior must be related to different microstructural features. As far as the alumina grain size is concerned, it shows little variation with the zirconia content. This variation might have an influence on the residual stresses [28], being this factor a second order factor on the aging sensitivity.

*3.4    AFM*

The same zone was successively observed by SEM in BEI mode and AFM in contact mode at different aging treatment times. Micrographs are shown in Fig 7. The microstructure is clearly visible, with dark grain boundaries (the brighter the image the higher the relief), and the grains were easily separated. When comparing AFM micrographs obtained after and before the aging treatment; it is consequently possible identifying grains that have transformed (arrows). Transformation induces a height increase, modifying the AFM contrast and consequently transformed grains brighten up. The microstructural environment of the grains can be identified on the SEM micrograph. From this figure, it can be observed that transformation occurs equally in isolated grains and agglomerates; agglomerates did not appear as preferential nucleation sites.

Self-accommodating martensitic variant pairs were also observed in some of the largest grains of the samples. An example is shown in Fig 8. The effect of thermal etching on the grains surface relief can be seen on the surrounding grains. Alumina is indeed more sensitive than zirconia to the etching treatment, and the surface relief is modified accordingly to its crystallographic orientation to the surface [23]. However, it is worth noticing the scale of thermal etching induced relief is much lower than that due to transformation. The rippled surface induced by the etching treatment shown on the bottom alumina grain in Fig 8 is limited to 5 nm, while the surface increase after transformation observed on the zirconia grain is about 60 nm. Cross sections of both the transformed grains and the alumina etching steps are given in Fig. 9. Again, the contrast of AFM derivate image is proportional to the rate of relief variation and not to the magnitude of the relief, which explains why the thermal etch steps in the alumina are apparently more impressive. The transformation-induced relief is clearly more important than the etch-induced relief.

## 4 DISCUSSION

### 4.1 Mehl-Avrami-Johnson (MAJ) model

**Comparison with 3Y-TZP**

For 3Y-TZP [25], the relationship between the amount of monoclinic phase and aging time at various temperatures can be described by a sigmoïdal law. This might be modeled by the MAJ equation, i.e.

$$f = 1 - \exp\left(-(bt)^n\right)$$

where f is the transformation fraction, t the time, b and n are constant dependent on the material. Moreover, b is a thermally activated parameter, providing the apparent activation energy Q:

$$b = b_0 \exp\left(-\frac{Q}{RT}\right)$$

where $b_0$ is a material constant, R the gas constant and T the temperature. This model is based on the nucleation and growth nature of the transformation. By plotting the transformation curve at different temperatures, it is possible measuring the apparent activation energy (Fig 10) and the exponent n (Fig 11). In the MAJ model, the n exponent is related to nucleation and growth conditions. A value between 3 and 4 will indicate a nucleation and 3D-growth behavior. This value is indeed observed for unconstrained zirconia (3Y-TZP), where no matrix prevents the transformation. In the present case, the overall behavior is different. For all compositions (i.e. up to 20 vol.% of 3Y-TZP), n is kept below 1, and varies linearly with the zirconia content. This would suggest that only nucleation occurs and the growth stage is not present. This might be easily interpreted considering the microstructure and the restricting influence of the matrix on the transformation. For low zirconia contents, the grains are distributed quite evenly in the matrix; the agglomerates fraction is quite low, so that the majority of the zirconia grains are isolated in the matrix. If one grain is transforming, the transformation will not propagate to surrounding zirconia grains, since alumina is found in-between. Near to near propagation of the transformation is therefore not possible. This is indeed what is observed on the AFM scans. When the zirconia fraction is increasing, zirconia grains get closer from each other and more agglomerates are found, as zirconia paths are building in the matrix. Above the percolation points, continuous paths of zirconia grains exist. The existence of a zirconia percolation threshold above which the transformation is propagating has been recently demonstrated [14]. Since percolation is a 3D phenomenon, only portion of these paths are visible in 2D observations (cross sections), explaining the apparent presence of zirconia agglomerates.

When a grain is transforming, triggering the transformation of a neighboring grain becomes more likely, as more and more neighboring zirconia grains are found, allowing the transmission of the transformation shear strain. There will be therefore a tendency toward the growth stage (propagation of the transformation). However, this phenomenon is clearly limited by the upper limit of zirconia content chosen in this study. Of all the materials chosen here, two compositions (17 and 20 vol.%) belong to the composites above the percolation point. Hence, only these materials could possibly show propagation of the transformation along the zirconia grains paths.

Finally, the low values observed for the n exponent are worth being discussed. A value of 1, according to the MAJ model [29], will refer to nucleation only. Lower values as observed here would therefore suggest that not only the growth stage is absent, but also the nucleation stage is not occurring as freely as it would in unconstrained zirconia. The alumina matrix does not only prevent the propagation of the transformation, but it does also partially prevent nucleation of monoclinic phase. This effect might be explained by the higher Young's modulus of alumina, compared to zirconia. When the zirconia content is increasing, nucleating the monoclinic phase becomes easier, as the apparent matrix modulus is reduced. This is very clearly observed in Fig. 11, where the n exponent is found to vary linearly with the zirconia content. The matrix has still an effect above the percolation threshold, explaining the low values measured for the n exponent.

**Activation energy: comparison with 3Y-TZP**

From the XRD experiments carried out here, it was possible measuring an apparent activation energy. The plot of ln(b) versus 1/T, providing this value, is shown in Fig 10. A value of 78 kJ/mol is extracted from the plot, with a good accuracy (correlation coefficient of more than 98%). This energy might be compared to the one previously measured (106 kJ/mol) for 3Y-TZP [25]. The first point to notice is that even if somewhat lower, the activation energy is in the same range of order. Several factors are to be considered before intending any strict comparison. Even if the concordance between the fitting curve and the experimental values in the Fig. 10 plot is quite good, it should be kept in mind that these values come from XRD experiments, which provide a limited resolution. The signal in 3Y-TZP is quite important and less subjected to experimental conditions variations. However, as far as ZTA are concerned, the signal is lower, since less zirconia is concerned with the X-ray emission, e.g. for A16TZY, only 16% of the sample at the most might provide a signal. For the first stage of nucleation, when 10 vol.% of the zirconia phase is transformed, this means that only 1.6 vol.% of the sample is at the origin of the signal. The signal noise influence becomes greater and measurements are less precise. It could therefore be concluded that is it quite difficult to compare precisely the precise values of the apparent activation energy. However, the values plotted in Fig. 10 seem to indicate a very low deviation from the theory (good fitting curve). With all the restrictions discussed above, we could propose that the difference of activation energy measurements between 3Y-TZP and ZTA may be related to the underlying mechanisms. The value of 106 kJ/mol corresponds to the nucleation and growth stages, while the 76 kJ/mol measured for ZTA here is related to nucleation only. There are no fundamental reason for the two stages (nucleation and growth) having the same activation energy. These experiments would therefore indicate lower activation energy for the nucleation stage. For the case of 3YTZP only, a growth stage (propagation of the transformation) is possible because the matrix is transformable. For 3YTZP-alumina composites, the matrix (alumina) is non-transformable (for low 3YTZP content), so that the growth stage is not possible. If the accuracy of the data due to low monoclinic concentration measurement is limited, the nucleation and growth nature of the transformation was validated by local SEM and AFM observations.

*4.2 Martensitic transformation and influencing microstructural parameters*

The t-m transformation has been the object of a very large number of studies, and the martensitic nature of the transformation is now commonly recognized. One of the most remarkable features of the martensitic transformation is the presence of self-accommodating variants in the transformed monoclinic phase. It was recently shown [20] possible observing these variants in 3Y-TZP by AFM. This type of relief is shown in Fig. 8, where some variants pairs, related to each other by a 90° angle (very similar to the ones of Fig. 2 in Ref. [20] , can be seen on the left hand side of the grain. Observing such variants in ZTA is quite difficult, considering the scale at which the transformation is occurring and zirconia grain size (<1μm). However, providing a very high quality probe is used, i.e. a radius of curvature as low as possible, observing some variants on the largest grains of the composite is possible. Not only does it validate the martensitic transformation in these materials, but also it is worth mentioning that only half of the grain is transformed. This grain was first observed after two hours of aging at 140°C, and other observations were made after 10 and 40 hours of aging. The right hand side of the grain remained untransformed even after 40 hours of aging. An extensive study has been dedicated to the observation of martensitic relief in zirconia-based ceramics [30]. Such behavior was never observed. The grains are either completely transformed after a long time or not at all. The hypothesis of a different orientation of variants pairs on the right hand side, leaving the surface undisturbed after transformation, is crystallographically forbidden. The partial transformation of the grain must therefore be considered. This behavior might be accounted for by the influence of the alumina matrix. As it can be seen on the SEM picture, one zirconia grain is lying on the left of the grain and another one just below. The constraining effect of the matrix is therefore reduced on the left hand side, which was therefore able to transform. However, due to the difference in Young's modulus, as described before, the matrix restricts severely the transformation [31]. On all the other sides of the grain, only alumina is found, so that the formation of variants of large size is more difficult.

The case of transformation of a zirconia grain in a similar situation but of smaller size is given in Fig. 12. As it can be seen from the SEM-BEI micrograph in insert, two small zirconia grains are lying side to side. The left hand side grain clearly transformed during the autoclave treatment performed between the two AFM micrographs, as the relief height increase is obvious. The situation is similar to that of Fig. 8, in that the grain of interest is surrounded by non-transformable alumina matrix, except on one side. However, the transformation strain accommodation is easier, due to the smaller grain size, so that the whole grain was able to transform. The relief increase of the other grain is probably a "pop-out" effect related to the transformation of the first grain. The constraining effect of the alumina matrix was not important enough to prevent the transformation as previously. The GBH is also too important to allow observing martensite variants.

## 5    CONCLUSIONS

The autoclave aging behavior of zirconia toughened alumina composites was studied by X-Ray diffraction, scanning electron microscopy and atomic force microscopy.

It was shown the transformation was occurring by a nucleation mechanism only, for low zirconia fractions, as compared to nucleation and growth classically observed in yttria-stabilized zirconia (3Y-TZP). The growth stage of the transformation is limited by the non-transformable matrix and occurs only when zirconia grains clustering is found. The time-temperature equivalency of the transformation was verified and an apparent activation energy of 78 kJ/mol was measured, value related for the nucleation stage only. The influence of the alumina matrix was assessed by local SEM and AFM observations. Not only does alumina limit the transformation rate, but also in certain cases, constraining effect of the matrix might be greater than the driving force for transformation once the transformation is initiated in a grain, leading to partially transformed grains. Transformation was observed to occur in both agglomerates and isolated zirconia grains. The microstructural environment appears therefore more important for the aging sensitivity than the grain size.

## ACKNOWLEDGEMENTS

Authors are indebted to the CLYME (Consortium Lyonnais de Microscopie Electronique) for using the XL30 ESEM-FEG and to the CLAMS (Consortium de Laboratoires d'Analyse par Microscopie à Sonde locale) for using the nanoscope. The authors would like to acknowledge the European Union for the financial support under the GROWTH2000, project BIOKER, reference GRD2-2000- 25039.

| 3Y-TZP vol.% | Alumina grain size (μm) | 3Y-TZP grain size (μm) |
|---|---|---|
| 10 | 1.30 ±0.02 | 0.51 ±0.02 |
| 13 | 1.30 ±0.03 | 0.54 ±0.03 |
| 17 | 1.20 ±0.03 | 0.47 ±0.03 |
| 20 | 1.20 ±0.02 | 0.50 ±0.02 |

Table 1: Alumina and zirconia grain size for the different compositions.

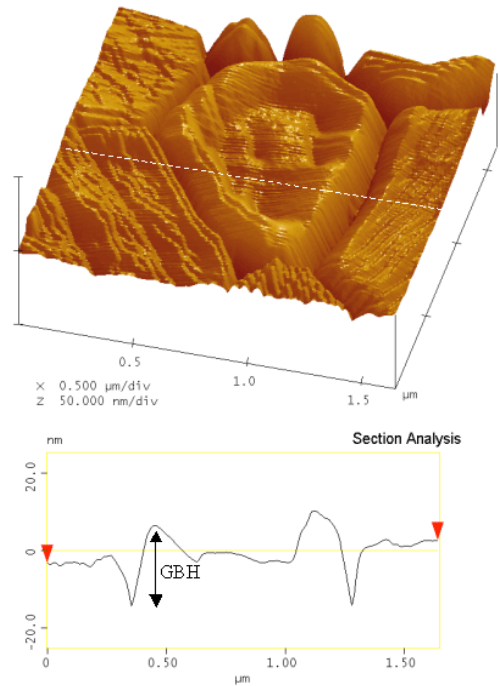

Fig. 1: AFM height image explaining the measurement of the grain boundary height, for a sample etched at 1723 K. The surface profile along the dashed line is plotted below.

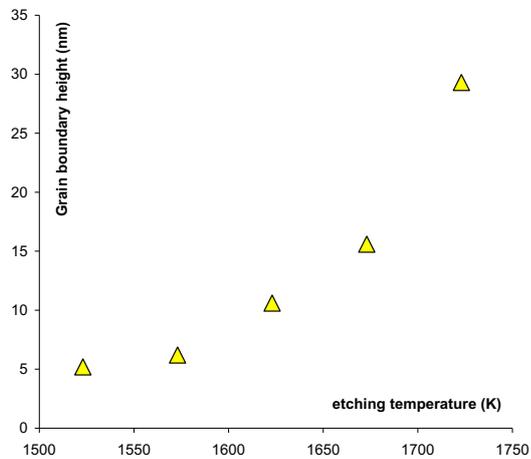

Fig. 2: Evolution of the average grain boundary height as a function of the etching temperature, for a 12 min treatment. Alumina with 17 vol.% (3Y-TZP).

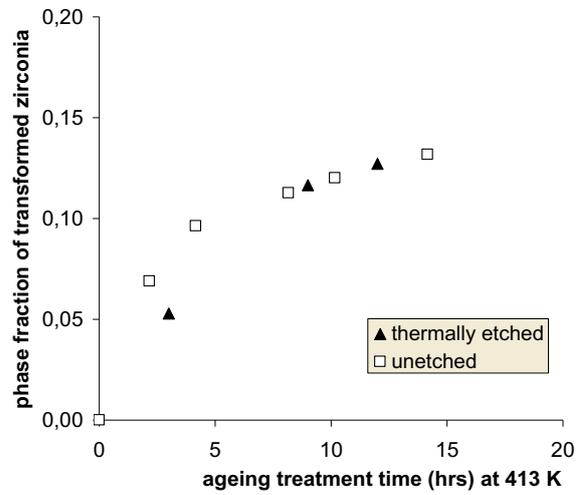

Fig. 3: XRD data showing the evolution of the monoclinic phase fraction as a function of aging treatment time, for samples unetched and thermally etched. No significant differences are observed. Alumina with 17 vol.% (3Y-TZP).

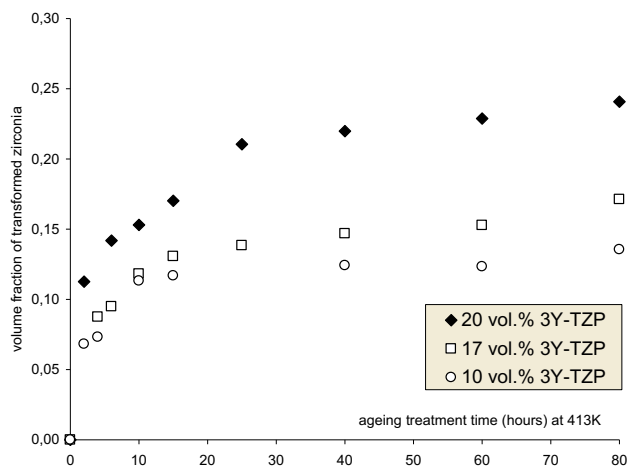

Fig. 4: XRD data showing the evolution of volume fraction of transformed zirconia as a function of the aging treatment time, for several composites (different 3Y-TZP content).

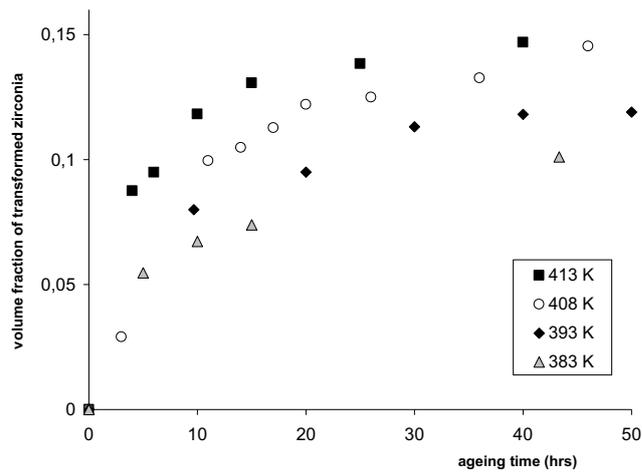

Fig. 5: XRD data showing the volume fraction of transformed zirconia as a function of the aging treatment time for different temperatures. Alumina with 17 vol.% (3Y-TZP).

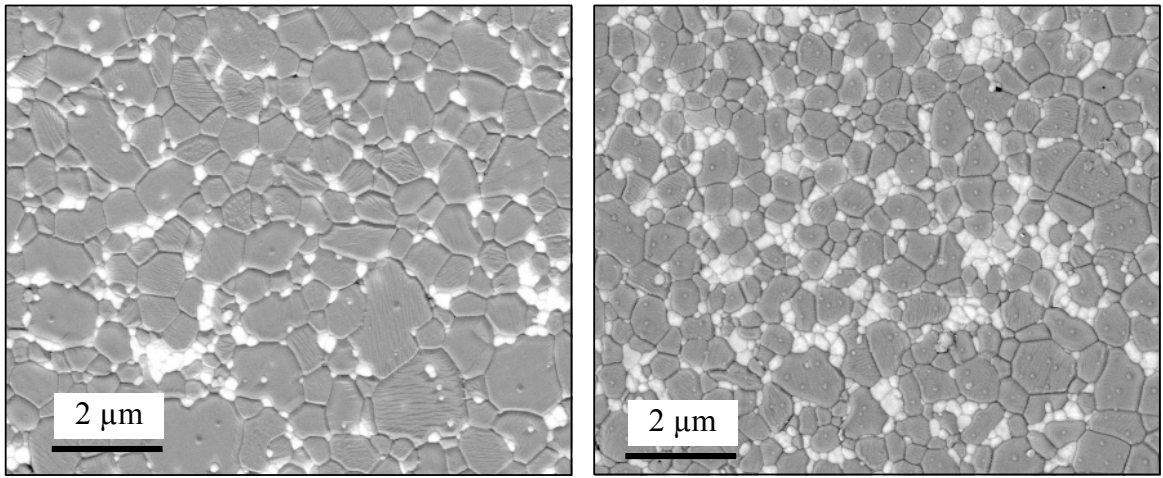

Fig. 6: SEM micrograph of the microstructures of two different compositions, i.e. 10 and 20 vol.% of (3Y-TZP).

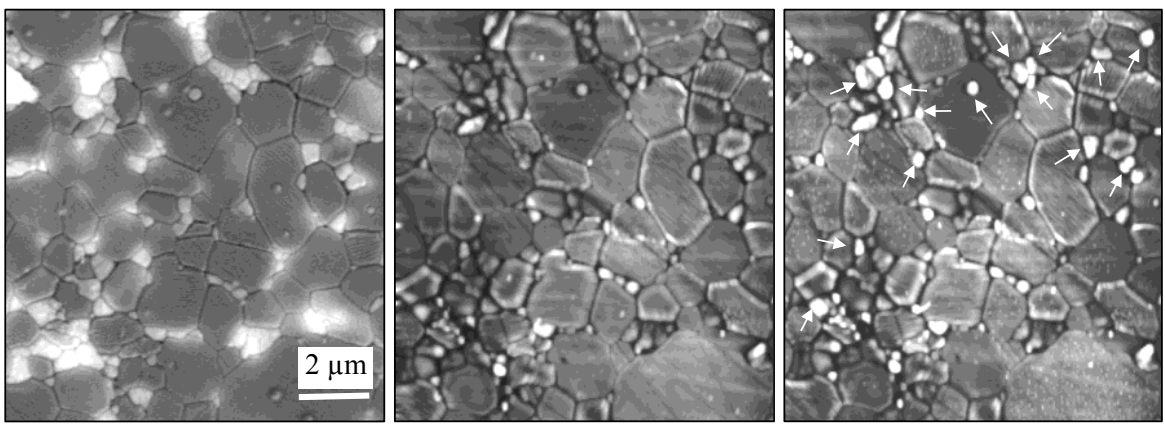

Fig. 7: SEM (left) and AFM (middle and right) observation of the transformation of the surface grains, at different aging treatment times (0 hrs and 5 hrs at 413 K), for the same zone of the sample (Al$_2$O$_3$-13 vol.%(3Y-TZP)). Arrows indicate the grains that transformed during the treatment stage between the two observations.

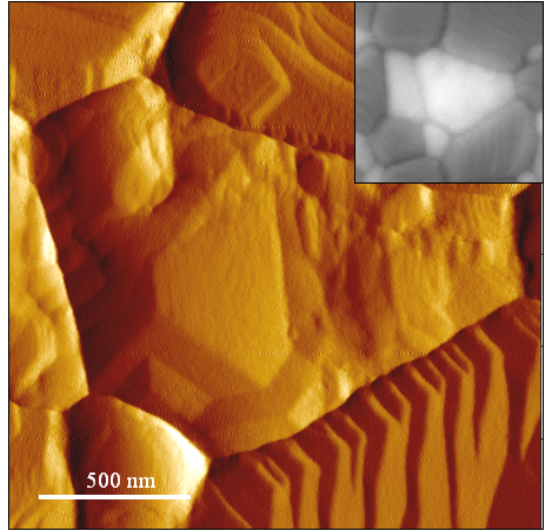

Fig. 8: AFM derivate image and SEM image of a partially transformed grain (40 hrs at 413 K in autoclave) in ($Al_2O_3$-17 vol.%(3Y-TZP)). Self accommodated martensitic variants are clearly visible, and the microstructural environment of the grain could be identified on the SEM micrograph (phase contrast, zirconia appears very bright).

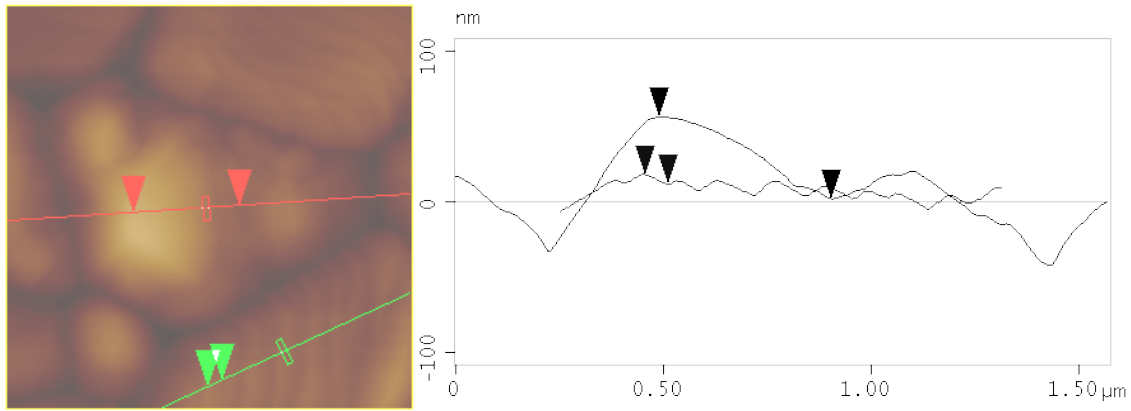

Fig. 9: Detail of Fig. 8. AFM height image and cross sections along the dashed lines for comparing the relative height variations of the martensitic relief (54 nm) and of the etching steps in alumina (6 nm).

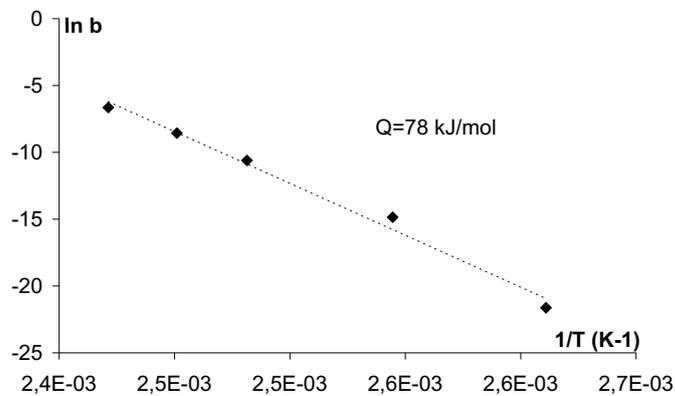

Fig. 10: Graph providing the apparent activation energy of the transformation, derived from XRD data's. See text for definition of the parameters. Alumina with 17 vol.% (3Y-TZP).

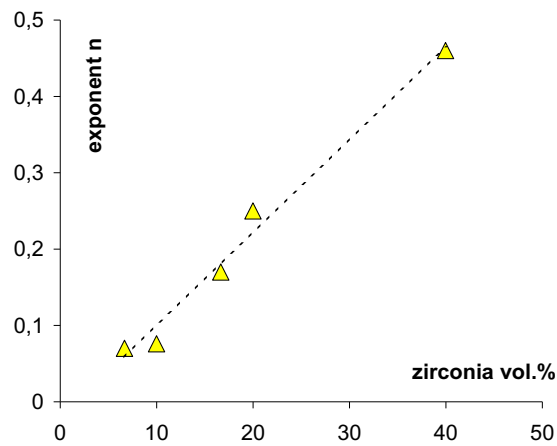

Fig. 11: Evolution of the n exponent as a function of the zirconia content in the composite (obtained from Fig. 4)

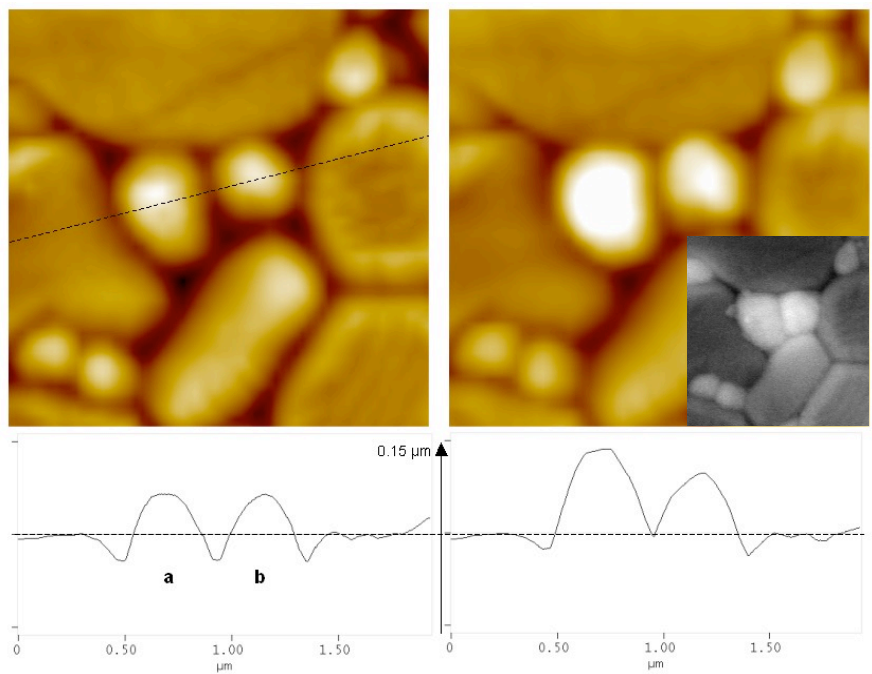

Fig. 12: AFM height image and cross sections along the dashed line, showing the transformation of a grain occurring during the autoclave treatment between the two micrographs ($Al_2O_3$-17 vol.%(3Y-TZP)). Nature of the various grains (3Y-TZP or $Al_2O_3$) can be identified by the SEM-BEI micrograph in insert.